# Spread of COVID-19: Adult Detention Facilities in LA County

Maria Barriga Beltran, Wendy Cano, Apichaya Chumsai, Haik Koyosan, Debbie Lemus, Sandra Tenorio, Jongwook Woo

FEMBA, California State University, Los Angeles

**Abstract:** We analyze the spread of COVID-19 cases within adult detention facilities in Los Angeles (LA) county. Throughout the analysis we review the data to explore the range of positive cases in each center and see the percentage of people who were positive for COVID-19 against the amount of people who were tested. Additionally, we see if there is any correlation between the surrounding community of each detention center and the number of positive cases in each center and explore the protocols in place at each detention center. We use the cloud visualization tool SAP Analytics Cloud (SAC) with the data from the California government website through adult detention facilities in LA County. We found that (1) the number of confirmed cases at the facilities and the surrounding communities are not related, (2) the data does not represent all positive cases at the facility, and (3) there are not enough tests at the facilities.

## 1. Introduction

With the assistance of SAP Analytics Cloud we use the information retrieved from the California Open Data Portal to look into COVID-19 cases in LA adult detention facilities [1, 7, 8]. We examine the information between July 31, 2020, and March 3, 2022. We also referred to the city of Los Angeles public health website to gather data around the amount of tests taken and the amount of positive cases. This will help us see whether positive COVID-19 cases were within a similar range between institutions or if there were differences, as well as look at the percentage of people getting tested throughout the specified date range. Additionally, this paper will look into the protocols in place at each of the institutions.

## 2. Related Work

In the midst of all the changes that took place after COVID-19 began spreading, there were numerous articles providing information around the impact to the community, businesses and people in different areas. There was constant information provided on a daily basis regarding positive cases and tests in communities. However, updates surrounding positive cases and tests in LA County detention centers were not readily provided [2, 3].

There was some research done around how COVID-19 impacted the day to day expectations in detention centers. People tried to gather information and they wrote articles about the situation, so we used this as an opportunity to try and further analyze this data a little closer to get an understanding of the situation taking place in these detention centers; however, the articles available were constantly contradicting each other. Some of the contradicting information we found stated that there weren't enough tests available to be distributed to detention centers, or that there were inmates who asked to get tested but they were denied [4 – 7].

## 3. Work Findings

In March of 2020, COVID-19 started spreading quickly throughout the county of Los Angeles. News and updates were consistently provided on a daily basis regarding positive cases and the amount of tests that were taken. However, updates surrounding positive cases and exams in LA County detention centers were not readily provided. We used this as an opportunity to analyze this data a little closer to get an understanding of the situation taking place in these detention centers.

### 3.1 Detention Centers and Positive Cases

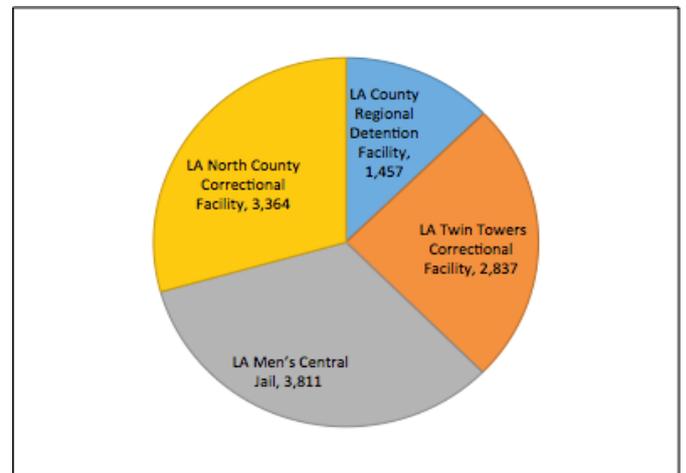

Figure 1a. Total number of inmates in each facility (July 31, 2020-March 3, 2022)

There are seven adult detention centers in LA County, but there was not enough data available to be able to do an analysis on three of them so we were unable to look into these three. The four detention centers we examined are the following: LA County Regional Detention Facility, LA Twin Towers Correctional Facility, LA Men's Central Jail, and LA North County Correctional Facility.

Between the dates of July 31, 2020, and March 3, 2022, the LA County Regional Detention Facility had the most cases at 487 positive cases. Their number of inmates was about 1,457, so they had a 33% positive rate. LA Twin Towers Correctional Facility had the second most positive cases at 428. They had about 2,837 inmates, so they had a 15% positive rate. LA Men's Central Jail had a 9% positive rate with 341 positive cases and 3,811 inmates, and LA North County Correctional Facility had a 3% positive rate with 103 positive cases and 3,364 inmates.

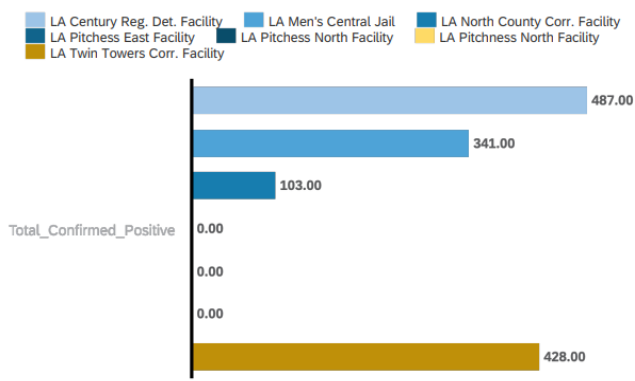

Figure 1b. Total confirmed positive cases in each detention facility examined (July 31, 2020-March 3, 2022)

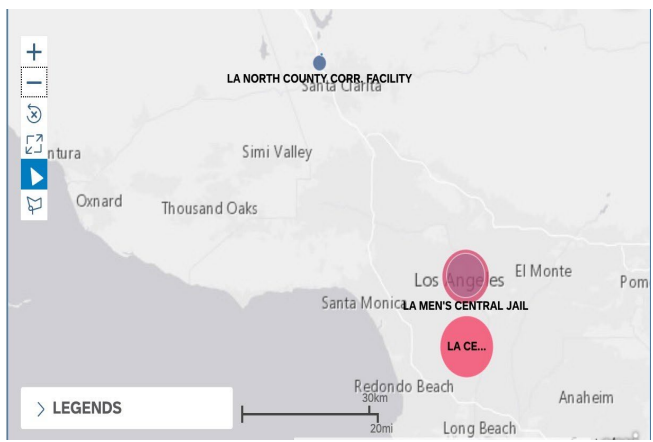

Figure 2: location of each facility in LA county with confirmed cases (July 31, 2020-March 3, 2022)

## 4. Background and Existing Work

The dataset we used was retrieved from the California Data Portal, its statewide open portal agency created to improve collaboration, expand transparency and lead to innovation. The data consists of weekly reports done by each county in California. We are focusing on the communities surrounding the detention centers we examined. The size of the data set is 43.3 KB and the following table shows specifications of what information was used.

Table 1: Data Specification

| Data Set | Size (Total) |
| --- | --- |
| Total_confirmed_positive | 43.3 KB |

### 4.1 Testing and Positive Rates in Surrounding Areas

In addition to looking at the testing rates and positive rates within each detention center, we also examined the testing rates and positive rates in the surrounding community of each detention center. The North County Correctional Facility is located in the Castaic area, LA Century Regional Detention Facility is in the Lynwood area, the Twin Towers Correctional Facility and the LA Men's Central Jail are located in the Chinatown area. We looked at the amount of people who tested positive against the amount who got a COVID-19 test. Below is a table that lists the numbers of these findings. Chinatown had the least amount of people testing positive. 20 out of 100 people who got tested received positive results.

Therefore, we expected to find that the LA Men's Central Jail and Twin Towers, which are both located in Chinatown, to have the lowest positive cases per 100 people. However, both of these detention centers had the highest positive rates with LA Men's Central jail at 5 out of 100 people and the Twin Towers facility at 6 out of 100 people. Even though these two detention centers have the highest positive cases per 100 people, they are still lower than the amount in its community. This displays a disparity between the surrounding community and the detention center, which brings up the question around protocols and procedures inside each center.

One might think the disparity in numbers is due to the facility having effective procedures and systems in place. However, we need to also explore the question around tests, such as whether tests were readily available to inmates and whether people with symptoms were accepting getting tested. If people were unable to get tested when they had symptoms or they refused to take a test, then the numbers we have would not be one hundred percent accurate of the situation at hand.

Table 2: Population in each community who got tested and were positive for COVID-19

| Community | Persons Tested | Positive Cases | Ratio (Positive vs. Tested) |
| --- | --- | --- | --- |
| Chinatown | 133,228 | 26,820 | 20 out of 100 people |
| Lynwood | 100,872 | 24,705 | 25 out of 100 people |
| Castaic | 68,562 | 25,906 | 38 out of 100 people |

Table 3: Inmate population in each detention center who got tested and were positive for COVID-19

| Detention Center (LA County) | Persons Tested | Positive Cases | Ratio (positive vs. tested) |
|---|---|---|---|
| North County Facility | 2,644 | 103 | 4 out of 100 people |
| Century Detention Facility | 11,965 | 487 | 4 out of 100 people |
| Men's Central Jail | 7,335 | 341 | 5 out of 100 people |
| Twin Towers | 7,410 | 428 | 6 out of 100 people |

### 4.2 Policies put into place at facilities

Once COVID-19 was confirmed in LA County's adult detention centers, there were numerous policies put into place to help control the spread between inmates. If anyone had flu-like symptoms they were immediately quarantined, and if they were in a dorm setting, then the entire dorm had to quarantine for 14 days to see if anyone else showed symptoms. Visitations were also put to a stop for about a year after COVID-19 infections began. Additionally, if someone showed flu-like symptoms or tested positive for COVID-19, then they were not allowed to go to court. Below is a list of additional steps taken at each detention center.

1. Dedicated housing areas became isolation and quarantine spaces.
2. Screened incoming inmates outside before entering the Inmate Reception Center (IRC) facility.
3. Inmates brought to the IRC who responded "yes' to the screening questions or displayed symptoms were masked and isolated in the onsite Correctional Treatment Center (CTC).
4. Newly arrived arrestees who were asymptomatic, but who reported close contact with infected people, were masked and placed in quarantine if appropriate.
5. Implemented additional cleaning, sanitation, and medical isolation procedures
6. Masks were provided.

### 5. Data Engineering

In order to properly analyze the data, a comma-separated values (CSV) file was downloaded from the California open data portal. The first step was to add coordinates (latitude and longitude), so that we could sort the data and create visualizations by location in SAP Analytics Cloud. We attempted to use the data in SAP Analytics Cloud without the coordinates but we weren't able to properly create a story and evaluate the data. Adding the coordinates to the file allowed us to see the quantity of positive cases by location in a graph. Having these visuals helped us continue with our evaluation. Once this was done we were able to upload the file to SAP.

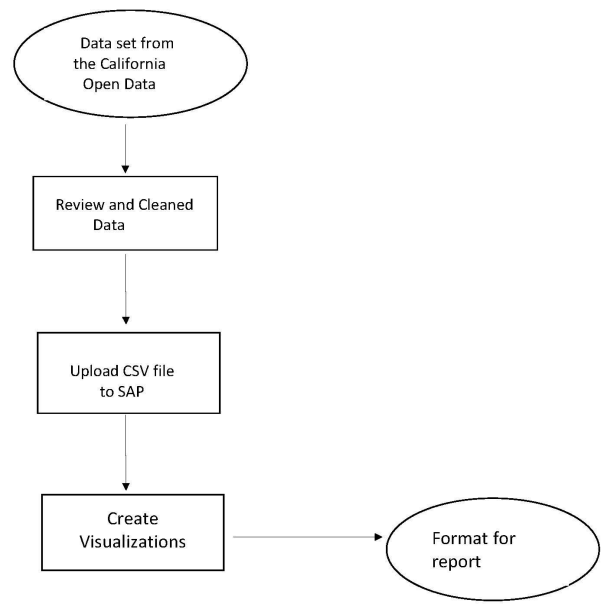

Figure 3: Implementation Flowchart

### 6. Data Analysis and Visualization

In trying to visualize the data, we quickly realized that we were trying to analyze more data than was manageable for one project. We initially attempted to review the data for adult detention centers in California and compare it with juvenile detention centers in California.

However, because of the volume of information, we decided to shift the focus of our research to positive COVID-19 cases in LA County detention centers. With this new approach we would look at the number of inmates in each center, positive cases, positive percentage rate, compare it to the community of each detention center, and the overall policies in place at the detention centers after the beginning of COVID-19. As we looked at this information we had to evaluate and decide how to transfer this information to charts and graphs.

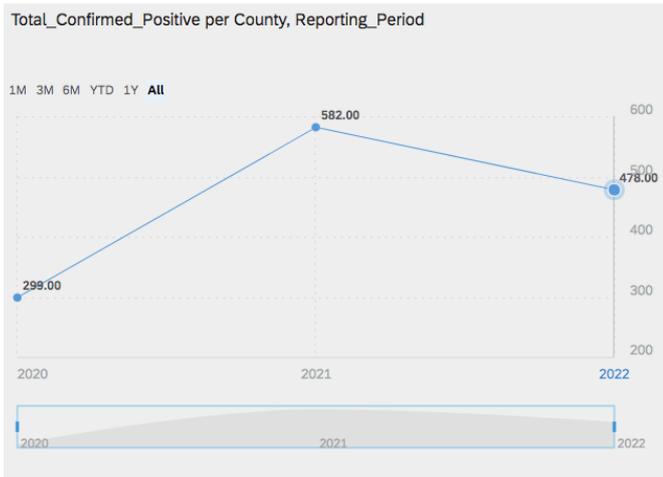

Figure 4: Time series of total confirmed positive cases in the four LA County detention centers.

## 7. Conclusion

As we analyzed the above data and information there are a few things that were identified, which include:

- The positive cases in these four detention centers do not provide an accurate representation of the situation compared to its surrounding community. The rate per 100 people was significantly lower than its community, and the community with the lowest positive ratio had the detention centers with the highest positive ratio.
- An accurate representation might not be possible since people who were showing symptoms and refused to take a COVID-19 test might have tested positive. This would mean that people who were positive would not be represented in the data.
- This also brings up the fact that there might not have been enough tests to distribute to each detention center for everyone who needed them.